\documentclass[10pt,conference]{IEEEtran}
\IEEEoverridecommandlockouts

\usepackage{amsmath,amssymb,amsfonts}
\usepackage{algorithmic}
\usepackage{graphicx}
\usepackage{textcomp}
\usepackage{xcolor}
\usepackage{enumitem}
\usepackage{comment}
\usepackage{balance}
\usepackage{enumitem}
\usepackage{multirow}
\usepackage{caption}
\usepackage[ruled,linesnumbered]{algorithm2e}
\usepackage{bm}
\usepackage{url}
\usepackage[colorinlistoftodos]{todonotes}
\usepackage{subfigure}
\usepackage{tcolorbox}
\usepackage{booktabs}
\usepackage{amssymb}
\usepackage{fontawesome5}
\usepackage[hidelinks]{hyperref}
\SetAlFnt{\small}
\SetAlCapFnt{\small}
\SetAlCapNameFnt{\small}
\usepackage{algorithmic}
\algsetup{linenosize=\small}
\SetKwInput{KwInput}{Input}                
\SetKwInput{KwOutput}{Output}              

\newboolean{showcomments}
\setboolean{showcomments}{true}
\ifthenelse{\boolean{showcomments}}
{ }

\ifthenelse{\boolean{showcomments}}
{ }

\ifthenelse{\boolean{showcomments}}
{ }

\ifthenelse{\boolean{showcomments}}
{ }

\def\BibTeX{{\rm B\kern-.05em{\sc i\kern-.025em b}\kern-.08em
    T\kern-.1667em\lower.7ex\hbox{E}\kern-.125emX}}

\begin{document}

\title{Revisiting Neuron Coverage Metrics and Quality of Deep Neural Networks}

\author{\IEEEauthorblockN{Zhou Yang, Jieke Shi, Muhammad Hilmi Asyrofi, David Lo}
\IEEEauthorblockA{\textit{School of Computing and Information System} \\
\textit{Singapore Management University}\\
Singapore \\
\{zyang, jiekeshi, mhilmia, davidlo\}@smu.edu.sg}
}

\maketitle

\begin{abstract}
Deep neural networks (DNN) have been widely applied in modern life, including critical domains like autonomous driving, making it essential to ensure the reliability and robustness of DNN-powered systems. As an analogy to {\em code coverage} metrics for testing conventional software, researchers have proposed {\em neuron coverage} metrics and {\em coverage-driven} methods to generate DNN test cases.
However, Yan et al. doubt the usefulness of existing coverage criteria in DNN testing. They show that a coverage-driven method is less effective than a {\em gradient-based} method in terms of both uncovering defects and improving model robustness.

In this paper, we conduct a replication study of the work by Yan et al. and extend the experiments for deeper analysis. A larger model and a dataset of higher resolution images are included to examine the generalizability of the results. We also extend the experiments with more test case generation techniques and adjust the process of improving model robustness to be closer to the practical life cycle of DNN development. Our experiment results confirm the conclusion from Yan et al. that coverage-driven methods are less effective than gradient-based methods.
Yan et al. find that using gradient-based methods to retrain cannot repair defects uncovered by coverage-driven methods. They attribute this to the fact that the two types of methods use different perturbation strategies: gradient-based methods perform differentiable transformations while coverage-driven methods can perform additional non-differentiable transformations. We test several hypotheses and further show that even coverage-driven methods are constrained only to perform differentiable transformations, the uncovered defects still cannot be repaired by adversarial training with gradient-based methods. Thus, defensive strategies for coverage-driven methods should be further studied.
\end{abstract}

\begin{IEEEkeywords}
Deep Learning Testing, Coverage-Driven Testing, Software Quality
\end{IEEEkeywords}

\section{Introduction}
\label{sec:intro}

Deep neural networks (DNN) models have achieved human-level performance on many tasks (e.g., image classification~\cite{ciregan2012multi}, speech recognition~\cite{abdel2014convolutional}, game playing~\cite{mnih2015human}, etc.), and have also been ubiquitously applied in our modern life, including autonomous driving~\cite{levinson2011towards} and so on. As a result, ensuring the reliability and robustness of systems that are powered by DNN techniques is of great importance, especially in critical domains like autonomous driving~\cite{levinson2011towards} and healthcare~\cite{choi2017gram}.

Researchers propose {\em code coverage} metrics to guide the testing of conventional software systems, like line coverage and branch coverage. To adequately test software systems, programmers are encouraged to collect test suites that have better coverage metrics~\cite{gopinath2014code}. 
Many tools are designed to automatically generate effective test cases by optimizing code coverage.
Unlike the conventional software whose logic is encoded in the control and data flows, DNN models, which are usually viewed as unexplainable black-boxes, encode their behaviors in the activations of millions (even trillions) of neurons. 
As an analogy to coverage-driven test cases generation for conventional software, researchers have proposed {\em neuron coverage} metrics to test DNN models. For example, Pei et al.~\cite{DeepXplore} define the neuron coverage of a test suite as the ratio of the number of unique activated neurons for all test inputs and the total number of neurons in a DNN model. 
Many studies designed various structural neuron coverage metrics to guide the testing of DNN models, e.g. DeepHunter~\cite{DeepHunter}, DeepGauge~\cite{DeepGauge}, etc. Such methods are called {\em coverage-driven} methods in the following parts of this paper, and the other branch of works that expose DNN defects by utilizing model parameters to computer gradients are called {\em gradient-based} methods.

However, some recent studies~\cite{li2019structural,FSE_Yan, harel2020neuron} suggest that the current neuron coverage metrics may not be effective enough to guide the generation of DNN test cases. Yan et al.~\cite{FSE_Yan} use a coverage-driven method (DeepHunter~\cite{DeepHunter}) to generate test cases and use these examples to retrain models. They find that when the coverage metrics increase, the quality metrics of models do not increase accordingly. They also compare the coverage-driven method with Projected Gradient Descent (PGD)~\cite{PGD}, a gradient-based method, finding that DeepHunter is less effective than PGD. Harel-Canada et al.~\cite{harel2020neuron} add a regularization term to the PGD method so it can generate examples with higher neuron coverage. But they find that this additional regularization term harms PGD's ability in generating adversarial examples.

Although prior works have shed new light into neuron coverage, they are conducted on relatively small datasets and models, with a limited number of gradient-based attacks, which motivates us to replicate their studies on other datasets and models, as well as extend the experiments with more test case generation methods to explore additional research questions. This extended replication study has the following features:

\begin{itemize}[leftmargin=*]
    \item \textbf{More datasets and models}: Prior studies on diverse software engineering tasks, e.g., automated debugging~\cite{10.1145/1101908.1101949,10.1145/2597008.2597148}, etc., have shown that findings can be different when data of different characteristics are used (e.g.,~\cite{10.1109/ICSE.2017.62,kim2021datasets}). This replication study is conducted on an additional dataset: EuroSAT~\cite{eurosat}, which has images of the larger resolution. We also analyse a larger model, ReseNet-56~\cite{resnet}. In total, we perform the experiments on 4 datasets and 9 models.
    \item \textbf{More gradient-based methods}: Yan et al.~\cite{FSE_Yan} only include PGD~\cite{PGD} as a representative of gradient-based methods. This replication study compares coverage-driven methods with an additional and computationally less expensive method, FGSM~\cite{FGSM}. 
    \item \textbf{More coverage-driven methods}: DeepHunter~\cite{DeepHunter}, the coverage-driven method used in~\cite{FSE_Yan}, performs a series of image transformations, which can be divided into two types: differentiable transformations (e.g., slightly changing pixel values) and non-differentiable transformations (e.g., rotation and blurring). We include two variants of DeepHunter, each of which is coverage-driven but can only perform one type of image transformations to generate test cases.
    \item \textbf{Detailed analysis on coverage-driven methods}: We perform deeper analysis of the test cases generated by coverage-driven methods. We adjust the process of improving model robustness to be closer to the practical life cycle of DNN models (iteratively generating test cases and retraining models) and then analyze how model quality metrics change with neuron coverage metrics. 
    We categorize test cases into two types: {\em coverage-improving examples} and {\em coverage-remaining examples}; the former can improve coverage metrics while the other cannot. We use the two types of test cases to retrain models and see whether there is a significant difference between retrained models' quality metrics. By introducing the two variants of DeepHunter, we also analyze whether defects uncovered by certain types of transformations are harder to be repaired by retraining with gradient-based methods.
\end{itemize}

Our extended experiments replicate conclusion made by the previous study~\cite{FSE_Yan} and also have some interesting observations. 
First of all, we confirm that the quality metrics do not monotonously change when models are retrained with datasets of increasing coverage metrics.
Second, we find that using coverage-improving examples to retrain models does not lead to better model quality than using coverage-remaining examples. We perform Mann-Whitney U test~\cite{mann1947test} on the model quality metrics of models retrained using the two types of examples, and the results suggest no significant difference, which further doubts the usefulness of existing coverage criteria in DNN testing.
Third, we replicate the observation that gradient-based methods are more effective than coverage-driven methods in uncovering defects. 

Besides, although coverage-driven methods generate fewer failed test cases, such uncovered failure cannot be repaired by adversarially training models with gradient-based methods. Similarly, using test cases generated by coverage-driven methods to retrain models also cannot repair defects uncovered by gradient-based methods. Prior works~\cite{FSE_Yan} attribute this phenomenon to the fact that the two types of methods use different perturbation strategies: gradient-based methods perform {\em differentiable} transformations like slightly modifying pixel values, while coverage-driven methods can use additional {\em non-differentiable} transformations like rotation and blurring. We test a hypothesis to show that even coverage-driven methods are constrained to only perform differentiable transformations, the uncovered defects still cannot be repaired by adversarial training with gradient-based methods, suggesting that more works on fixing coverage-driven test cases are required. 


The rest of this paper is organized as follows. Section \ref{sec:background} briefly describes background information. Section \ref{sec:metrics} discusses the neuron coverage and model quality metrics. We explain the 4 research questions in Section \ref{sec:rqs} and report the results in Section \ref{sec:experiment}. Section \ref{sec:related_work} surveys related works and Section \ref{sec:conclusion} concludes this replication study.
\section{Background}
\label{sec:background}
This section introduces some background information, including the concepts and definitions of deep neural networks, adversarial attack and adversarial training.

\subsection{Deep Neural Networks}

This paper focuses on deep neural networks (DNN) for single label classification. A DNN classifier can be formalized as a function $f: X \to Y$, mapping a set of inputs $X$ into a set of labels $Y$. The output of the DNN classifier is a probability distribution $P(Y|x)$, which is the probability that an input vector $x \in X$ belongs to each class of labels in $Y$, and the label of the input $x$ is deemed as the class with the highest probability. A DNN classifier $f$ usually contains an input layer, a number of hidden layers, and an output layer; each layer consists of many {\em neurons}. The parameters $\theta$ of the DNN are the weights of each connected edge between neurons of two adjacent layers. For an input vector $x$, the output of DNN $f_\theta(x)$ can be computed as the weighted sum of the outputs of all the neurons. Given a training dataset $D=\{(x_i, y_i)\}_{i=1}^N$ containing $N$ input examples $\{x_1, x_2, \dots, x_N\}$ and the corresponding ground-truth labels $\{y_1, y_2, \dots, y_N\}$, a DNN classifier learns to optimize the following objective:
\begin{equation}
    \min_\theta \frac{1}{N} \sum^N_{i=1} \mathcal{L} (f_\theta(x_i), y_i)
\end{equation}
where $\mathcal{L}$ is the loss function to calculate the penalties for incorrect classifications. The DNN are trained by minimizing the loss on the training set and updating the parameters $\theta$ accordingly.

\subsection{DNN Robustness}

Numerous work have shown that DNN models are not robust to adversarial perturbations in the input data. Adding some human-imperceptible perturbations $\delta$ to an image $x$ can fool a DNN classifier to produce a wrong output, which can be formalized as:
\begin{equation}
    f(x + \delta) \neq f(x) 
\end{equation}
The input with perturbations $x+\delta$ is referred to as an {\em adversarial example} or a {\em failed test case}, and the process of imposing perturbations to the input data is referred to as the {\em adversarial attack}. Utilizing the gradient information of DNN models to generate such small perturbations is shown to be effective. We investigate the following two {\em gradient-based} methods in our paper.

\vspace{0.15cm}
\noindent{\bf FGSM.} Fast Gradient Sign Method (FGSM)~\cite{FGSM} is the earliest adversarial attack that utilizes the gradient information. Given a DNN model $f$ and an input $x$, FGSM first calculates the gradient $\nabla_x$ of the loss function with respect to the input $x$, and then adds a small perturbation $\epsilon$ following the direction of gradients on the input $x$ to generate adversarial example $x_{adv}$. The idea of FGSM can be formalized as follows.
\begin{equation}
    x_{adv} = x + \epsilon * sign(\nabla_x \mathcal{L} (f(x), y))
\end{equation}
where $\mathcal{L}$ is the loss function, $x$ is the original input data and $y$ is the original label. The perturbation is scaled by a constant factor $\epsilon$ (to ensure that the perturbation is small enough) and added to the input data.

\vspace{0.15cm}
\noindent{\bf PGD.} Projected Gradient Descent (PGD)~\cite{PGD}, which is based on FGSM, attempts to find the adversarial examples using a multistep scheme. In each step, given a DNN model $f$ and an input $x$, it first adds a small random perturbation within a given bound, and then performs FGSM to apply the gradient information on the perturbation of the input. The process continues until successfully generating an adversarial example or time-out. 

The above gradient-based methods adopt one {\em differentiable} transformation, i.e., directly changing pixel values. However, they cannot perform {\em non-differentiable} transformations, e.g., rotation and blurring. The coverage-driven methods are free to use different types of image transformations, which we explain in Section \ref{sec:rqs}.

\subsection{Adversarial Training}

When software systems contain bugs, programmers check the program logic to fix the bug. DNN models define a data-driven programming paradigm where models learn logic by updating parameters automatically to minimize the loss on training datasets. Inspired by this feature, the key idea of adversarial repair is to improve DNN models' robustness by augmenting the training dataset with adversarial examples and perform retraining. The objective of retraining this model is to minimize the loss on the augmented dataset instead of only on the original dataset, which can be formalized using the following equation.
\begin{equation}
    f^{*} = \mathop{\arg\min}_{f} \mathop{\mathbb{E}}\limits_{\substack{(x, y_{true}) \\ \sim  D_{a} \cup D}}[\mathcal{L}(f(x), y_{true})]
\end{equation}
where $D$ is the original training data and $D_{a}$ is the generated adversarial examples. 
The achieved robustness highly depends on the adversarial examples used, thus the re-trained model can mainly defend the specific adversarial attacks and may not be robust against other attacks~\cite{madry2018towards, AdversarialTraining}. In this work, we also investigate whether the gradient-based adversarial training methods can defend the adversarial examples generated by coverage-driven methods.

\section{Coverage Metrics and Model Quality Metrics}
\label{sec:metrics}
In this section, we formally describe the DNN coverage metrics and model quality metrics used in this paper. For more detailed information on these metrics, we refer interested readers to the following papers~\cite{DeepXplore, DeepHunter,FSE_Yan}. The implementation of these metrics can be found in our replication package~\cite{replication}.

\subsection{DNN Coverage Metrics}

For a neuron $n$ in a DNN model $f$, we view it as an \emph{activated neuron} if given an input $x$, output of a neuron $n$ is larger than a threshold $t$. The set of activated neurons in the DNN is defined as:
\begin{equation}
    AC(x, t) = \{n|f_\theta(n,x) > t \}
\end{equation}

Pei et al.~\cite{DeepXplore} define a model's neuron coverage given a set of inputs $T$ as follows:
\begin{equation}
    NC(T, t) = \frac{|\{n| \exists x \in T, f_\theta(n,x) > t \}|}{|N|}
    \label{math:NC_wrong}
\end{equation}


Xie et al.~\cite{DeepHunter} define several structural coverage metrics and present a fuzz testing framework for DNN. These metrics are defined as follows:

\noindent{\bf $k$-multisection Neuron Coverage (KMNC):}
For a neuron $n$, the upper and lower boundary of its output values on training data can be denoted as $\text{up}_{n}$ and $\text{low}_{n}$, respectively. The $k$-multisection neuron coverage measures how thoroughly the boundary $[\text{low}_{n}, \text{up}_{n}]$ are covered by the given test inputs $T$. To quantify KMNC, the boundary $[\text{low}_{n}, \text{up}_{n}]$ is divided into $k$ equal sections (i.e., $k$-multisections), and $S^n_m$ denotes the $m$-th section. Then $\phi(x,n) \in S^n_m$ means the $m$-th section is covered by at least one input data $x \in T$. The $k$-multisection neuron coverage is defined as:
\begin{equation}
    KMNC = \frac{\sum_{n \in N} | \{ S^n_m | \exists x \in T : \phi(x, n) \in S^n_m \}}{k \times |N|}
\end{equation}

\noindent{\bf Neuron Boundary Coverage (NBC):}
The boundary range $[\text{low}_{n}, \text{up}_{n}]$ is derived from training data. For new test inputs $T$, the output values of neurons may fall into $(-\infty, \text{low}_{n})$ or $(\text{up}_{n}, +\infty)$ instead of the above boundary range. The range $(-\infty, \text{low}_{n}) \cup (\text{up}_{n}, +\infty)$ is referred as the corner case regions of neurons, and the \emph{Upper Neuron Coverage} (UCN) and \emph{Lower Neuron Coverage} (LCN) are defined as follows:
\begin{equation}
    \begin{aligned}
    U C N &=\left\{n \in N \mid \exists x \in T: f_\theta(x, n) \in\left(\text{high}_{n},+\infty\right)\right\} \\
    L C N &=\left\{n \in N \mid \exists x \in T: f_\theta(x, n) \in\left(-\infty, \text{low}_{n}\right)\right\}
    \end{aligned}
\end{equation}

For the input test data $T$, NBC measures to what extent the corner-case regions are covered outside the boundary derived from training data. The definition of NBC is as follows:
\begin{equation}
    NBC = \frac{|UCN| + |LCN|}{2 \times |N|}
\end{equation}

\noindent{\bf Strong Neuron Activation Coverage (SNAC):} 
Similar to NBC, strong neuron activation coverage measures how many upper-corner neurons are covered by the given test inputs $T$.
The definition of SNAC is the ratio of the number of \emph{Upper Neuron Coverage} (UCN) and the total number of neurons in the DNN:
\begin{equation}
    SNAC = \frac{|UCN|}{|N|}
\end{equation}

\noindent{\bf Top-$k$ Neuron Coverage (TKNC):}
TKNC is a layer-level coverage testing criterion, which measures the ratio of neurons that have at least been the most active $k$ neurons of each layer on a given test set $T$ once. The definition of TKNC is as follows:
\begin{equation}
    TKNC = \frac{|\cup_{x \in T}(\cup_{1 \leq l \leq L} top_k(x,l))|}{|N|}
\end{equation}
where $l$ denotes a layer in the DNN, and $top_k(x,l)$ denotes the neurons which have largest $k$ output values of layer $l$ on the given test data $T$.

\vspace{0.1cm}
\noindent{\bf Top-$k$ Neuron Patterns (TKNP)}: 
TKNP measures how many patterns of the top-$k$ neurons are covered by the given test data $T$. For a test input $x$, the top-$k$ neurons of each layer forms a specific pattern, which is an element of $k_1 \times k_2 \times \dots \times k_l$. The $k_l$ is a subset of top-$k$ neurons among all neurons in the $l$-th layer. The definition of TKNP is as follows:
\begin{equation}
    TKNP = |\{(top_k(x,1)), ..., (top_k(x,L)) | x \in T|\}|
\end{equation}

\subsection{DNN Model Quality Metrics}

The evaluation metrics of DNN model quality are centered around adversarial examples. The most common evaluation on model quality is to measure the model accuracy in the presence of adversarial examples. High accuracy under adversarial attacks means that the model is robust. There exists some metrics related to the model accuracy:

\vspace{0.1cm}
\noindent{\bf Misclassification Ratio (MR):}
\begin{equation}
    MR = \frac{1}{N} \sum^N_{i=1} \#(f(x^a_i) \neq y_i)
\end{equation}
where $x^a_i$ denotes adversarial examples and $N$ is the total number of them. $\#(f(x^a_i) \neq y_i)$ is the number of misclassified examples. Basically, MR calculates the percentage of misclassified examples over the whole input set. A DNN model with high quality should have a low MR.

\vspace{0.1cm}
\noindent{\bf Average Confidence of Adversarial Class (ACAC):}
\begin{equation}
    ACAC = \frac{1}{n} \sum^n_{i=1} p(x^a_i)_{f(x^a_i)}
\end{equation}
where $n$ is the total number of misclassified adversarial examples. And $p(x^a_i)_{f(x_i^a)}$ is the confidence towards the incorrect prediction $f(x^a_i)$. In general, ACAC measures the average confidence towards incorrect labels of adversarial examples, a high quality DNN model should have a low ACAC.

\vspace{0.1cm}
\noindent{\bf Average Confidence of True Class (ACTC):}
\begin{equation}
    ACTC = \frac{1}{n} \sum^n_{i=1} p(x^a_i) y_i
\end{equation}
Similar with ACAC, $n$ is the total number of misclassified adversarial examples, and $p(x^a_i)_{y_i}$ is the confidence towards the true labels of adversarial examples $y_i$. ACTC measures the prediction confidence of true labels of adversarial examples, a high quality DNN model should have a high ACTC.

In addition to the metrics about model accuracy, some metrics are related to the imperceptibility of adversarial examples. The intuition behind these metrics is that some adversarial examples are generated by injecting substantial perturbations into the original data. These examples can violate the implicit forms of realistic input data and DNN models can be easily fooled. But in such cases, the failures of DNN models are not necessarily due to the adversarial examples. The metrics related to the imperceptibility of adversarial examples are:

\noindent{\bf Average $L_p$ Distortion ($ALD_p$):}
\begin{equation}
    ALD_p = \frac{1}{n} \sum^n_{i=1} \frac{|| x^a_i - x_i ||}{|| x_i ||_p}
\end{equation}
where $||.||_p$ is the $\ell_p$ norm to measure the distortion between the perturbed image and the original example. More specifically, we use three kinds of norm distance to calculate to what extent the attacks perturb original examples when generating adversarial examples. $\ell_0$ calculates the number of pixels changed in adversarial examples, $\ell_2$ calculates the Euclidean distance between original examples and adversarial examples, and $\ell_{\infty}$ calculates the maximum change in all dimensions of adversarial examples. $ALD_p$ measures the average normalized $\ell_p$ distortion for misclassified adversarial examples. The adversarial example with high imperceptibility has small $ALD_p$.

\noindent{\bf Average Structural Similarity (ASS)}: 
\begin{equation}
    ASS = \frac{1}{n} \sum^n_{i=1} SSIM(x^a_i,x_i)
\end{equation}
ASS measures the average similarity between misclassified adversarial examples and their corresponding original examples. It utilizes a metric called $SSIM$ to quantify the similarity between two images~\cite{SSIM}. The adversarial example with high imperceptibility has larger ASS.

\noindent{\bf Perturbation Sensitivity Distance (PSD)}: 
\begin{equation}
    PSD = \frac{1}{n} \sum^n_{i=1} \sum^m_{j=1} x_{i,j} Sen(R(x_{i,j}))
\end{equation}
PSD evaluates human perception of adversarial perturbations by measuring difference between the changed pixel and its surrounding square region. For the $i$-th adversarial example, $m$ is the total number of pixels that example has and $x_{i,j}$ denotes the $j$-th pixel. $R(x_{i,j})$ represents the surrounding square region of $x_{i,j}$. $Sen(R(x_{i,j})) = 1/std(R(x_{i,j}))$, in which $std(R(x_{i,j}))$ denoting the standard deviation function. The adversarial example with high imperceptibility should have small PSD.

Another series of evaluation metrics indirectly reflect the model quality by measuring adversarial example robustness. Intuitively, if an adversarial example that is not robust (i.e., not always lead DNNs to make misclassifications), it cannot indicate low model quality when it is misclassified. We use the following metrics to evaluate adversarial example robustness in particular:

\noindent{\bf Noise Tolerance Estimation (NTE)}: 
\begin{equation}
    NTE = \frac{1}{n} \sum^n_{i=1} [p(x^a_i)_{f(x^a_i)} - max\{ p(x^a_i)_j \}]
\end{equation}
where $n$ is the total number of misclassified adversarial examples. $p(x^a_i)_{f(x^a_i)}$ is the confidence towards the incorrect prediction $f(x^a_i)$. $max\{ p(x^a_i) \}$ is the max confidence of all other labels. NTE estimates the noise tolerance of an adversarial example by measuring how much noise it can tolerate while keeping its misclassified label unchanged. A robust adversarial example should have a large NTE.

\noindent{\bf Robustness to Gaussian Blur (RGB)}: 
\begin{equation}
    RGB = \frac{\#(f(GB(x^a_i)) \neq y_i)}{\#(f(x^a_i) \neq y_i)}
\end{equation}
Here $GB$ denotes the Gaussian blur function which can reduce noises in images. $\#(f(GB(x^a_i)) \neq y_i)$ is the number of adversarial examples can maintain misclassified labels after Gaussian blur, while $\#(f(x^a_i) \neq y_i)$ is the total number of misclassified adversarial examples. The greater the RGB is, the more robust adversarial examples are in the input data.

\noindent{\bf Robustness to Image Compression (RIC)}: 
\begin{equation}
    RIC = \frac{\#(f(IC(x^a_i)) \neq y_i)}{\#(f(x^a_i) \neq y_i)}
\end{equation}
Similar with RGB, RIC counts how many adversarial examples can maintain their misclassified labels after being compressed. $\#(f(IC(x^a_i)) \neq y_i)$ denotes the number of adversarial examples can keep misclassified labels unchanged after Gaussian blur and $\#(f(x^a_i) \neq y_i)$ is the total number of misclassified adversarial examples. The greater the RIC, the more robust the adversarial examples.

\section{Research Questions}
\label{sec:rqs}
This section describes some research questions that explore whether neuron coverage metrics are effective criteria to help uncover defects in DNN models and improve the model quality. RQ1, 3, and 4 are also included in Yan et al.'s study~\cite{FSE_Yan}, and we extend them with additional analysis. RQ2 is newly designed for analyzing whether coverage-improving examples are more effective than coverage-remaining examples.

  

\subsection*{RQ1. Do model quality metrics improve when coverage metrics increase?}

Some research works suggest that developers should design test suites that can adequately cover more lines or branches of code to reveal more faults in the software and improve software quality by repairing these identified faults~\cite{google_coverage}. But researchers also conducted empirical studies to show that code coverage may not be a strong indicator of the test suite effectiveness when testing conventional software systems~\cite{icse_no_correaltion}. This research question explores how model quality metrics change when neuron coverage metrics are increasing.

To answer this research question, we adopt the same experiment setting as Yan et al.~\cite{FSE_Yan}. Based on the description in the paper~\cite{FSE_Yan} and the code in the corresponding replication package~\cite{CovTesting}, we introduce the process of how Yan et al. apply a coverage-driven method and how they measure the relationship between coverage metrics and model quality metrics\footnote{We use different notations as Yan et al.~\cite{FSE_Yan} but convey the same meaning. We have also consulted with the first author to get additional details.}. We assume that $M_0$ is the original model under test, $M_0$ is trained on the dataset $D_0$ and $T$ is the original test set for evaluating model performance. 
Firstly, $D_0$ is split into $k$ subsets, denoted by $\{d_1 \cdots d_k\}$. 
For each subset $d_i$, a coverage-driven method, which randomly performs some pre-defined image transformations (e.g., rotation, blurring, etc.) , is applied to generate a set of perturbed images $d'_i$. Each image in $d'_i$ is a tuple $\langle x,y\rangle$, where $x$ is an input image and $y$ is the label of this image. 

We explain the process of selecting the {\em coverage-improving examples} from $d'_1$. For each image in $d'_1$, if it is misclassified by $M_0$, we add it into $D_0$ and see whether the augmented dataset has higher neuron coverage metrics. We select all the images that can increase coverage metrics (i.e., they cover additional neurons) and augment them into $D_0$ to get $D_1$. Similarly, $D_i$ can be obtained by mutating $d_i$ and augmenting $D_{i-1}$. So $D_i$ is a superset of $D_{i-1}$, and the former covers more (at least no less) neurons in $M_0$ than the latter.
After recording the coverage metrics of each dataset in $\{D_1, \cdots, D_k\}$ on the model $M_0$, we use the $k$ datasets to retrain $M_0$ and obtain $\{M_1, \cdots, M_k\}$.
Then, we evaluate the quality metrics (described in Section \ref{sec:background}) of the $k$ retrained models and analyze how model quality metrics change when neuron coverage metrics are increasing.



  \begin{algorithm}[t]
    \caption{The refined process of iteratively generating test cases and improve model quality.} 
    \label{algo:new_rq1}
    \SetAlgoLined
    \KwInput{$M_0$: the original model under test, $D$: the original training dataset, $k$: times to split the training dataset, $t$: the test data to evaluate model quality metrics}
    \KwOutput{Coverage metrics and model quality metrics}
    \# split the training dataset into $k$ subsets $d_1 \cdots d_k$\;
    $\bm{d} = split(D_0, k)$\;
    \# generate examples that can improve coverage\;
    \For(){
      $i$ {\em in} $1 \cdots k$
      }{
          $D_{i} = D_{i-1}$\;
          \# randomly transform images in $d_k$\;
          $d'_k = mutate(d_i)$\;
          \For(){
            $\langle x,y\rangle$ {\em in} $d'_k$
          }{
            \If{ $y \neq M_{i-1}(x) \wedge nc(M_{i-1}, D \cup \langle x,y\rangle) > nc(M_{i-1}, D)$}{
                \# select coverage-improving examples\;
                $D_{i}.append(\langle x,y\rangle)$\;
             }
          }
        $M_{i} = retrain(M_{i-1}, D_i)$ \# change 2 \;
        \# measure coverage and model quality metrics\;
        $coverage_i = measure\_c(M_{i-1}, D_i)$ \# change 3 \;
        $quality_i = measure\_q(M_i, t)$\;
    }
  
  \algorithmicreturn{ $\bm{coverage}, \bm{quality}$}
  \end{algorithm}

\subsection*{RQ2. Are coverage-improving examples more effective in improving model quality?}

The process in RQ1 incrementally builds multiple test suites that differ in the ability in covering neurons and uses them to obtain different models by retraining the same original model. Yan et al.~\cite{FSE_Yan} claim that such a process is analogous to how the developers enhance their test suite over time to improve coverage. It is indeed the case when developers try to design better test suites for a single version of the software.

However, AI systems (as well as conventional software systems) are continuously evolving. Developers find and fix faults in the current version of the software and then deliver a new version. Before giving a new version of the software, regression testing is usually performed to see whether new changes break some original functions. Then, developers design test suites that can adequately test the {\em new} software. Such an iterative process is repeated to enhance software quality. We refine the retraining process in RQ1 to be closer to the aforementioned processes of developing DNN models, which is illustrated in Algorithm \ref{algo:new_rq1}.

In Algorithm \ref{algo:new_rq1}, we first split $D_0$ into $k$ subsets $\{d_k \cdots d_k\}$ (Line 2). For each subset $d_i$, some pre-defined image transformations (e.g., rotation, blurring, etc.) are randomly applied to generate the perturbed images $d'_i$ (Line 7). Then, we iterate all the images $\langle x,y\rangle \in d'_i$ (Line 8), and see whether a perturbed image is misclassified by the \textbf{current state} of model $M_{i-1}$ and can activate more neurons in $M_{i-1}$ \textbf{instead of the original model} $M_0$ (Line 9). If the condition in Line 9 is satisfied, we call such images {\em coverage-improving examples} and add them into $D_i$ for later usage. Then, $D_i$ is obtained by augmenting all the coverage-improving examples into the previous training dataset $D_i$, and $M_i$ is obtained by retraining $M_{i-1}$ on $D_i$ (Line 15). Then, we measure how well $D_i$ covers the original model $M_{i-1}$ and the model quality of the newly obtained model $M_i$ (Line 16 to 17). 

The above revised process can better depict how DNN models are iteratively tested and improved using the coverage-improving examples. In this research question, we also explore the values of {\em coverage-remaining examples}, which we define as the generated test cases that cannot cover additional neurons. The intuition is that if test cases with higher coverage are of better quality, the models retrained on such coverage-improving examples should be better than models retrained on coverage-remaining examples. We explore this research question by testing the following hypothesis:

\begin{tcolorbox}
  \textbf{Hypothesis 1}: Quality metrics of models trained with coverage-improving examples are significantly better than that of models trained with coverage-remaining examples.
\end{tcolorbox}

To test the hypothesis, we run Algorithm \ref{algo:new_rq1} to collect the coverage-improving examples, as well as run a modified version to collect and use the coverage-remaining examples. More specifically, we replace the coverage metrics-related condition in Line 9 with $y \neq M_{i-1}(x) \wedge nc(M_{i-1}, D _0\cup \langle x,y\rangle) \leq nc(M_{i-1}, D_0)$. Then, we compare the model quality metrics of models retrained on the two types of examples.

\subsection*{RQ3. Are gradient-based methods more effective in uncovering failure?}
Software testing is thought of as an effective way to uncover bugs in software systems. In this research question, we compare coverage-driven test case generation methods with two gradient-based methods (i.e., PGD~\cite{PGD} and FGSM~\cite{FGSM}) to see whether we can replicate the conclusion from Yan et al.'s experiment: PGD is more effective in uncovering failure in DNN models. We validate this conclusion by testing the following hypothesis:

\begin{tcolorbox}
  \textbf{Hypothesis 2}: Misclassification rates on test cases generated by gradient-based methods are significantly larger than that on test cases generated by coverage-driven methods.
\end{tcolorbox}

On top of the original experiment, we collect more data points with the following extension to experiments. The original experiment only selects PGD as the representative of gradient-based methods. PGD is effective but also expensive to use; it utilizes gradient information and performs attacks iteratively. We extend the experiment with another popular and computationally cheaper method, FGSM~\cite{FGSM}. Besides, we evaluate these test case generation methods on an additional model (i.e., ResNet-56~\cite{resnet}) and dataset (i.e., EuroSAT~\cite{eurosat}). 

Yan et al.~\cite{FSE_Yan} select DeepHunter~\cite{DeepHunter} as the single representative of coverage-driven methods, which pre-defines some image transformations like adjusting brightness, rotation, blurring, etc. 
The image transformations that DeepHunter applies can be divided into two types: differentiable transformations (e.g., slightly changing pixel values) and non-differentiable transformations (e.g., rotation and blurring). 
The non-differentiable transformations cannot be adopted by the gradient-based methods. We wonder whether the reason why coverage-driven methods are less effective is that the non-differentiable transformations it uses (e.g.,., rotation) may hurt the ability in uncovering defects. 
So we include the two variants of DeepHunter: DP-D and DP-N. The former is constrained to only perform differentiable transformations, and the latter can only perform non-differentiable transformations (more details of them will be given in Section \ref{subsec:settings}). We analyze the difference between two types of image transformations via testing Hypothesis 3:

\begin{tcolorbox}
  \textbf{Hypothesis 3}: Misclassification rates on test cases generated by DP-D are significantly larger than that on test cases generated by DP-N.
\end{tcolorbox}

\subsection*{RQ4. Can adversarial training repair failed test cases generated by coverage-driven methods?}
After failed test cases are discovered, programmers debug to find and repair faults in the programs. DNN models are developed in a data-driven paradigm: datasets are fed into DNN models, and models can automatically update parameters to learn patterns in the data, which can be viewed as an analogy to the automatic program repair techniques~\cite{ye2021automated}. 
As we mentioned earlier, after developing a new version of the software, regression testing is usually performed to see whether new changes break some original functions. Similarly, we evaluate the retrained models on the original benign datasets and test the following hypothesis:

\begin{tcolorbox}
  \textbf{Hypothesis 4}: The accuracy of retrained models on benign datasets is higher than accuracy of original models on the same datasets.
\end{tcolorbox}

Then, we explore whether gradient-based adversarial training can repair the defects uncovered by the coverage-driven test cases generation techniques. Yan et al.~\cite{FSE_Yan} find that PGD-based retraining can improve model performance against PGD attacks but cannot improve model accuracy against test cases generated by DeepHunter. We re-investigate this conclusion the conclusion by testing the Hypothesis 5 on more datasets and models.

\begin{tcolorbox}
  \textbf{Hypothesis 5}: Retraining models using gradient-based methods harms the model accuracy on test cases generated by coverage-driven methods.
\end{tcolorbox}

Beyond attempting to replicate the finding, we are also interested in whether the additional non-differentiable image transformations reveals defects that cannot be repaired by retraining with gradient-based methods, which use differentiable transformations only.
So we include the two variants of DeepHunter mentioned in RQ3 (DP-D and DP-N) and test another hypothesis:

\begin{tcolorbox}
  \textbf{Hypothesis 6}: After retaining using gradient-based methods, model accuracy on test cases generated by DP-D is higher than accuracy on test cases generated by DP-N.
\end{tcolorbox}

\section{Experiment Settings and Result Analysis}
\label{sec:experiment}
In this section, we first explain the experiment settings, including datasets, models under test, and hyper-parameter settings for test case generation methods. Then, we answer the research questions in Section \ref{sec:rqs} with detailed analysis.

\subsection{Datasets and Models under Test}
\begin{table}[]
    \centering
    \caption{Statistics of models under test and their performance on different datasets.}
    \begin{tabular}{ccccc}
      \hline
      Datasets                      & Models    & Layers & Parameters & Accuracy \\ \hline
      \multirow{3}{*}{MNIST~\cite{mnist}} & LeNet-1   & 9      & 3,246      & 98.67\%  \\
                                    & LeNet-4   & 10     & 25,010     & 98.65\%  \\
                                    & LeNet-5   & 11     & 44,426     & 98.93\%  \\ \hline
      \multirow{3}{*}{CIFAR-10~\cite{CIFAR10}}     & VGG-16    & 56     & 33,687,922 & 92.80\%  \\
                                    & ResNet-20 & 71     & 274,442    & 91.74\%  \\
                                    & ResNet-56 & 200    & 860,026    & 92.76\%  \\ \hline
      \multirow{3}{*}{SVHN~\cite{svhn}}         & SADL-1    & 19     & 6,531,786  & 89.70\%  \\
                                    & SADL-2    & 15     & 1,646,794  & 87.72\%  \\
                                    & SADL-3    & 33     & 2,923,050  & 92.57\%  \\ \hline
      \multirow{3}{*}{EuroSAT~\cite{eurosat}}      & VGG-16 &  56     & 33,687,922    &     96.50\% \\
                                    & ResNet-20 & 71     & 274,442    &     97.11\%     \\
                                    & ResNet-32 & 200    & 860,026    &    97.22\%    \\

                                    & ResNet-56 & 71     & 274,442    &    93.65\%      \\ \hline
      \end{tabular}
      \label{tab:statistics}
    \end{table}

Our experiments cover the three datasets investigated in Yan et al.'s study~\cite{FSE_Yan}: MNIST~\cite{mnist}, CIFAR-10~\cite{CIFAR10}, SVHN~\cite{svhn}, which are widely used in both coverage-driven and gradient-based test case generation for DNN models~\cite{DeepHunter, DeepGauge, DeepConcolic,DeepGini}.
The MNIST~\cite{mnist} dataset contain images of handwritten digits (from $0$ to $9$). It has $60,000$ $28 \times 28$ grayscale training images and a test set of $10,000$ images. The CIFAR-10~\cite{CIFAR10} dataset consists of 60,000 32x32 colour images that are categorized into 10 classes, with 6,000 images per class. We split the CIFAR dataset into training and testing sets with $50,000$ and $10,000$ images, respectively. We also include SVHN~\cite{svhn}, a real-world image dataset for recognizing house numbers obtained from Google Street View images. It has $73,257$ 32x32 colour training images and $26,032$ testing images. 
Beyond these datasets, we also include a more recently proposed dataset, EuroSAT~\cite{eurosat}, which contains satellite images covering 13 spectral bands and consisting of 10 classes with $27,000$ labelled and geo-referenced samples. Images in the EuroSAT dataset have 4 times higher resolution than the above three datasets and can be an appropriate benchmark to see whether the results from~\cite{FSE_Yan} can generalize to larger images.

As shown in Table \ref{tab:statistics}, each dataset corresponds to several baseline models. Considering that models used in the original experiment~\cite{FSE_Yan} are relatively small, whose coverage metrics can be easily improved, we investigate a larger model (i.e., ResNet-56~\cite{resnet}) in this replication study. We follow the instruction in a popular computer vision repository to train ResNet-56 on CIFAR-10 and EuroSAT datasets. Other baselines models are obtained from previous works~\cite{DeepHunter}. The statistics of these models (e.g., number of layers and parameters\footnote{The numbers of layers and parameters information is obtained using Keras APIs.}), as well as their performance on different datasets, are displayed in the Table \ref{tab:statistics}. All the datasets, models and results of the experiments can be accessed from our replication package~\cite{replication}.

\subsection{Test Case Generation and Parameter Settings} 
\label{subsec:settings}
We explain the test case generation methods we use in the experiments, as well as the hyper-parameter settings.
DeepHunter~\cite{DeepHunter} can perform 9 different image transformations. According to whether they are differentiable, they are divided into two groups. Four differentiable transformations include adjusting image contrast, brightness, changing some pixel values and adding noises. Five non-differentiable transformations include image translation, scale, shear, rotation and blurring. We build two variants of DeepHunter: DH-D, which can only perform differentiable transformations, and DH-N, which can only perform non-differentiable transformations.

According to the parameter settings in Yan et al.'s experiments~\cite{FSE_Yan}, we set the magnitude of the two gradient-based methods (the maximum perturbation allowed to each pixel) as MNIST: $0.3$, CIFAR-10: $\frac{16}{255}$, SVHN: $\frac{8}{255}$, EuroSAT: $\frac{16}{255}$. To generate test cases and perform adversarial training, we use a popular library: Adversarial Robustness Toolbox (ART) v1.8~\cite{art}.
To faciliate the large-scale experiments, we parallelly train and test DNN models on two servers that both have Intel(R) Xeon(R) CPU E5-2698 v4 @ 2.20GHz, 500 GB memory, and 8 NVIDIA GeForce P100. 
\subsection{Result Analysis}
\label{sec:result}

\subsubsection{Answers to RQ1}

\begin{figure}[t!]
    \centering
    \includegraphics[width=0.9\linewidth]{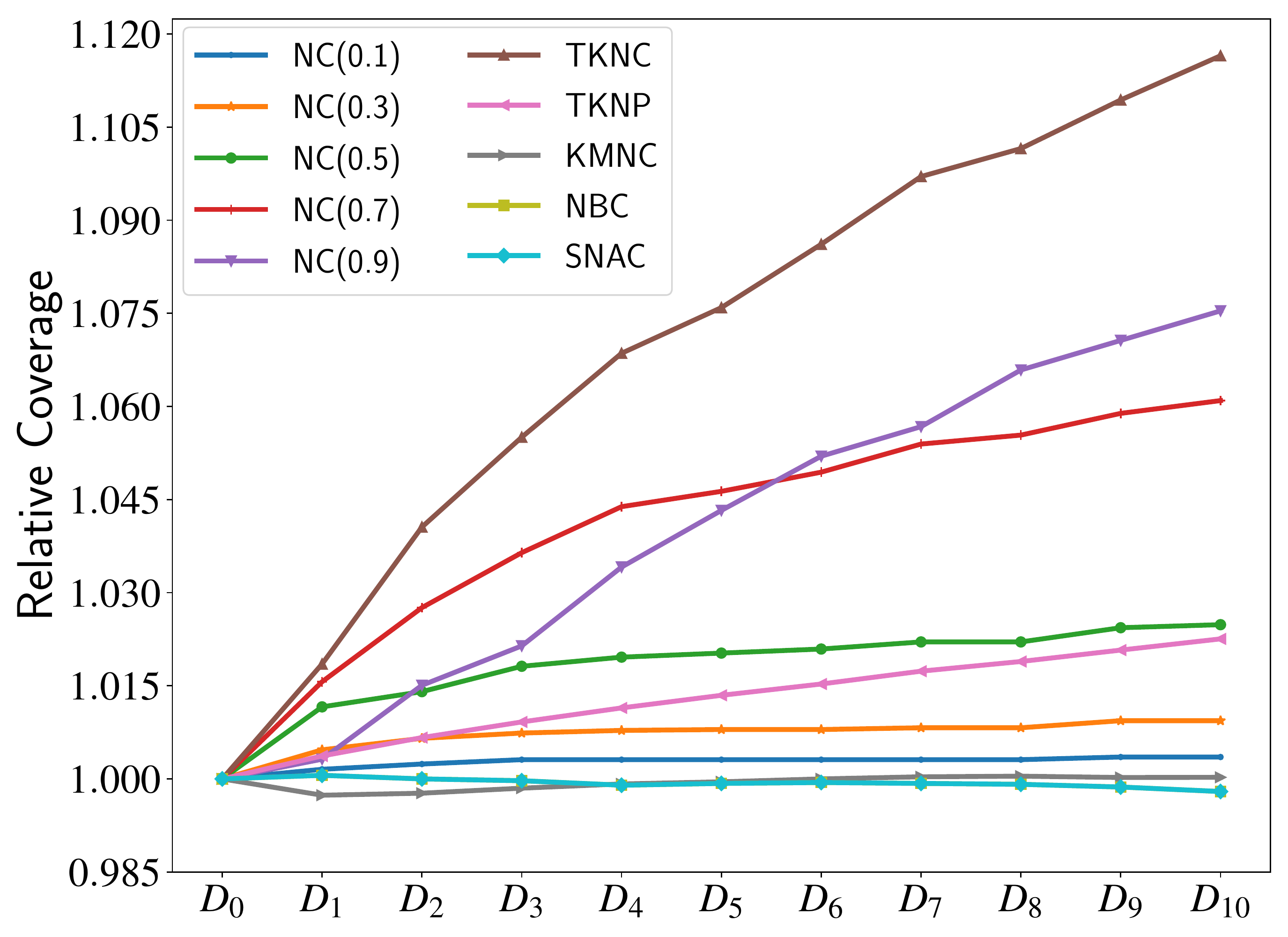}
    \caption{Trends of coverage metrics on the same original model.}
    \vspace{-0.5cm}
    \label{fig:rq1_coverage}
\end{figure}

\begin{figure}[t!]
    \centering
    \includegraphics[width=0.9\linewidth]{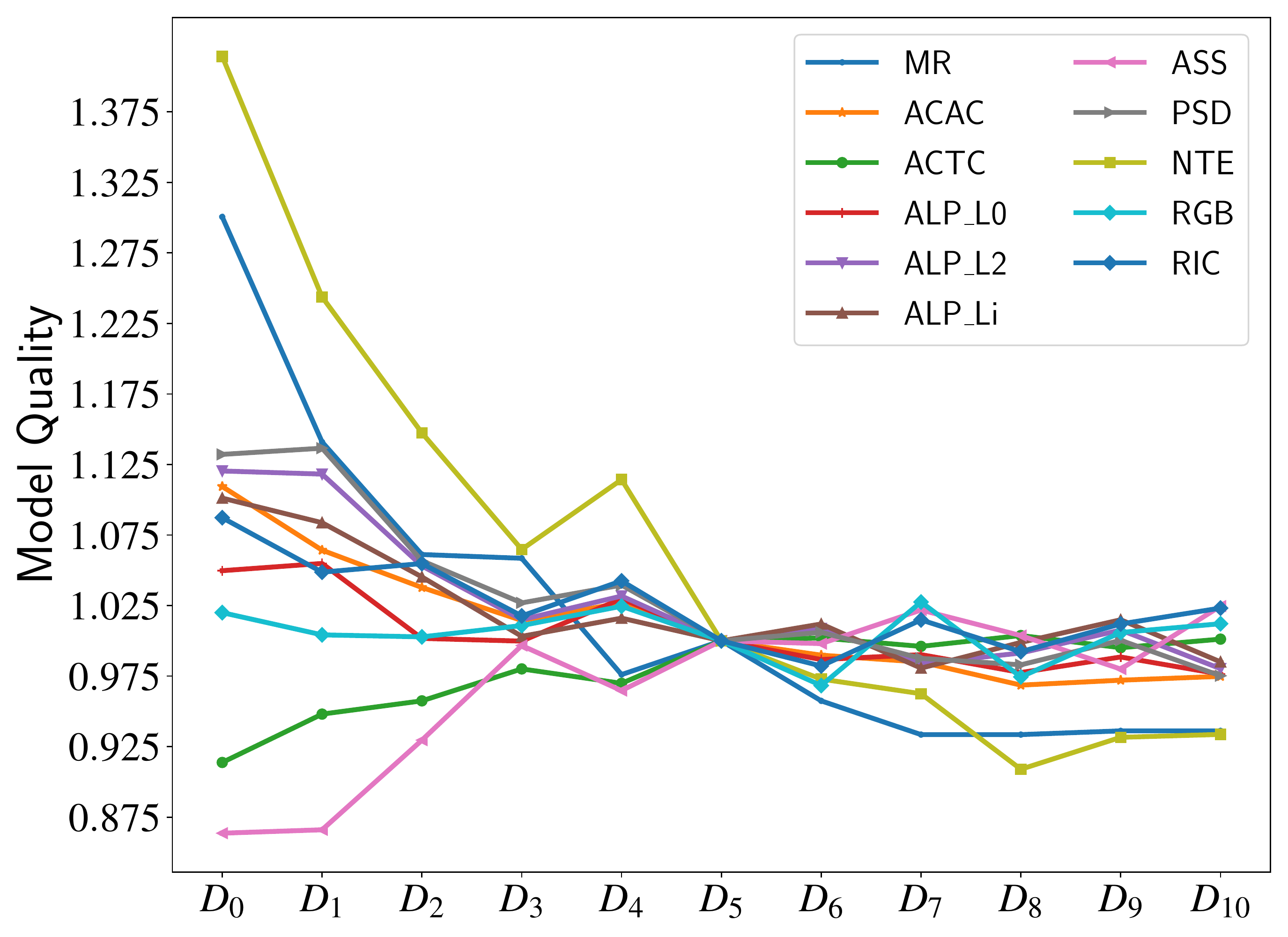}
    \caption{Trends of model quality metrics of models retrained using datasets with increasing coverage metrics.}
    \vspace{-0.5cm}
    \label{fig:lenet1-accuracy}
\end{figure}


According to the process described in Section \ref{sec:rqs}, for one model and its corresponding training data, we apply DeepHunter on the training data and select examples that can cover additional neurons in the model. Then we gradually augment these coverage-improving examples into the original training set to construct $k$ new datasets with different coverage metrics.
Fig \ref{fig:rq1_coverage} illustrates how well DNN models are covered by their corresponding augmented training data during this incremental process\footnote{Due to the limited space, we only visualize the results for retraining LeNet-1 on the MNIST dataset. The results for other models and datasets are available in our replication package \cite{replication}.}. For more clear visualization, all the metrics are normalized by being divided by the values at $D_0$.


We use these augmented data to retrain the original model and measure model quality metrics of these newly obtained models. Fig. \ref{fig:lenet1-accuracy} visualizes the trends of model quality metrics, and all the metrics are normalized by being divided by the values at $D_5$. We replicate Yan et al's observation that \textbf{none} of the model quality metrics is monotonous. So our answer to this research question is:

\begin{tcolorbox}
    \textbf{Answer to RQ1}: The quality metrics of models do not improve as the coverage metrics increase.
\end{tcolorbox}

\subsubsection{Answers to RQ2}

Unlike the experiments in RQ1, to answer RQ2, we intentionally select the examples that cannot cover additional neurons to retrain models. Such examples are called coverage-remaining examples because if we add them into the training set, the augmented training set has the same coverage metrics on the model.



\begin{table*}[t!]
    \centering
    \caption{Statistical tests for quality metrics of models trained on coverage-improving and coverage-remaining examples.}
    \begin{tabular}{@{}cccccccccccc@{}}
    \toprule
    \textbf{Robustness Metrics} & MR      & ACAC    & ACTC  & \begin{tabular}[c]{@{}l@{}}$\text{ALD}_{\ell_0}$\end{tabular} & \begin{tabular}[c]{@{}l@{}}$\text{ALD}_{\ell_2}$\end{tabular} & \begin{tabular}[c]{@{}l@{}}$\text{ALD}_{\ell_p}$\end{tabular} & ASS   & PSD     & NTE     & RGB   & RIC     \\ \midrule
    \textbf{$p$-value}                   & 0.245   & 0.100   & 0.487 & 0.079                                                            & 0.139                                                            & 0.188                                                            & 0.206 & 0.139   & 0.079   & 0.139 & 0.245   \\
    \bottomrule
    \end{tabular}
    \label{tab:rq2_metrics}
    \end{table*}

If a group of examples that cover more neurons is better, we expect that models retrained using coverage-improving examples should demonstrate better performance than models retrained using coverage-remaining examples. We measure the model quality metrics of models retrained using the two types of examples. We use the Mann–Whitney U test~\cite{mann1947test} to quantify the difference between the model quality of these two models. Table~\ref{tab:rq2_metrics} shows our statistical results. We can see that models trained using coverage-improving examples lead \textbf{no significant} improvement on the models trained using coverage-remaining examples ($p > 0.05$ for all robustness metrics). 
Therefore, we reject Hypothesis 1 and our answer to RQ2 is as follows.

\begin{tcolorbox}
    \textbf{Answer to RQ2}: Coverage-improving examples are not more effective in improving model quality.
\end{tcolorbox}

\subsubsection{Answers to RQ3}
Other than using the DeepHunter \cite{DeepHunter} and PGD \cite{PGD} from the original paper \cite{FSE_Yan}, we introduce three additional test case generation methods for further evaluation. The new gradient-based method we include is FGSM \cite{FGSM}, which is computationally cheaper than PGD. We split the image transformations that DeepHunter can operate into two types: differentiable transformations and non-differentiable transformations. More specifically, directly changing pixels (e.g., add noise, adjust brightness, etc.) is differentiable, while transformations like rotation and blurring are non-differentiable. We then introduce two variants of DeepHunter, each of which is coverage-driven but can only perform either differentiable or non-differentiable image transformations. We apply the five test case generation methods on the original testing datasets, and evaluate the models on the newly generated test cases. 

We collect the misclassification rates on test cases generated by both coverage-driven and gradient-based methods. By performing statistical testing on Hypothesis 2, we find that misclassification rates on test cases generated by gradient-based methods are \textbf{significantly higher} than that on test cases generated by coverage-driven methods, with a $p$-value less than $0.001$, which confirms the original observations.
Then, we compare the effectiveness of the two variants of DeepHunter: DP-D and DP-N. Our statistical test results reject the Hypothesis 3 as the $p$-value is $0.45$, much larger than $0.05$. It indicates that only using non-differentiable or differentiable transformations \textbf{does not significantly affect} the performance of coverage-driven methods. So our answer to the RQ3 is as follows.

\begin{tcolorbox}
    \textbf{Answer to RQ3}: Gradient-based methods are more effective in uncovering DNN failures ($p < 0.001$). Besides, differentiable and non-differentiable transformations do not show significant difference in terms of generating failed test cases as $p$-value is $0.45$.
\end{tcolorbox}

\subsubsection{Answers to RQ4}
In the process of developing conventional software systems, when failed test cases are identified, programmers or automated program repair tools need to localize the faults that cause the failed test cases and repair them to make test cases pass. Unlike the above test-driven development, the training of DNN models is a data-driven paradigm. To repair defects in DNN, we can augment the failed test cases (or examples of a similar distribution) into the training set and then retrain the models. 

In this research question, we mainly care about whether using gradient-based methods to perform adversarial training can repair defects uncovered by coverage-driven methods. For DeepHunter and its two variants, we first apply them on the training images, augment all the generated images into the original training set, and then retrain the models. For the two gradient-based methods, we use a popular library~\cite{art} to perform adversarial training.

We first test the hypothesis about whether adversarial training can harm model performance on the benign dataset. We find that on the benign datasets, the accuracy of retrained models (either using coverage-driven or gradient-based methods) is \textbf{statistically significantly smaller} than the original models ($p$-value is $0.002 < 0.05$). Although this conclusion is not explicitly mentioned by Yan et al., we can also observe similar results reported in tables from their paper \cite{FSE_Yan}.

Yan et al. \cite{FSE_Yan} make an observation that adversarial training using gradient-based methods has a limited impact on increasing model robustness on test cases generated by coverage-driven methods and even hurts the robustness in some cases. We test Hypothesis 5 to see whether we can confirm this observation. We first retrain models using gradient-based methods and independently evaluate obtained models on test cases generated by coverage-driven methods. A statistical test is performed to examine whether the accuracy of retrained models on coverage-driven test cases is lower than that of the original models. The statistical test results show a $p$-value of $0.024 < 0.05$, which suggests accepting Hypothesis 5. In other words, we \textbf{confirm} the above observation made in \cite{FSE_Yan}, which highlights that more works on defending coverage-driven methods are needed.

To explore whether the non-differentiable image transformations are the ones that cannot be repaired, we consider two variants of DeepHunter in our experiment, each of which is coverage-driven but can only perform either differentiable or non-differentiable image transformations. We test Hypothesis 6 and find that using gradient-based methods to repair models does not help increase more robustness against differential transformations than non-differentiable transformations. It suggests that coverage-driven and gradient-based methods uncover different defects in DNN models even using the same image transformations. Although coverage-driven methods can generate fewer failed test cases than gradient-based methods, these test cases can uncover different defects and more effective defensive approaches need to be designed specially.
Based on the above results, our answer to this research question is: 

\begin{tcolorbox}
    \textbf{Answer to RQ4}: Statistical test results with $p$-value~$< 0.05$ show that adversarial training with gradient-based methods cannot defend against coverage-driven test cases. 
\end{tcolorbox}

\subsection{Summary for Replication Results}
Table \ref{tab:summary} summarizes the hypotheses and the corresponding findings obtained by testing these hypotheses. A `\checkmark' means that a finding is explicitly made in Yan et al.'s paper~\cite{FSE_Yan}.

\begin{table}[]
    \caption{Summarization of replication results. Column 1 describe the findings that we get from testing hypotheses, column 2 describes hypotheses in Section \ref{sec:rqs}, and column 3 describes whether the findings are reported in \cite{FSE_Yan}.}
    \begin{tabular}{lcc}
    \hline
     Findings  & Hypothesis                                                                                            & \cite{FSE_Yan}     \\ \hline
     \begin{tabular}[c]{@{}l@{}}Coverage-improving examples are not more \\ effective than coverage-remaining examples.\end{tabular} ~  & H1 &   \\ \hline
    \begin{tabular}[c]{@{}l@{}}Gradient-based methods are more effective  \\ than coverage-driven methods in uncovering \\ defects.\end{tabular} & H2 &  \checkmark    \\ \hline
    \begin{tabular}[c]{@{}l@{}}Both non-differentiable and differentiable \\ transformations have similar ability in \\ uncovering defects.\end{tabular} & H3 &      \\ \hline
    \begin{tabular}[c]{@{}l@{}}Retraining models with generated test cases \\ can harm performance on benign datasets.\end{tabular} & H4 &  \checkmark    \\ \hline
    \begin{tabular}[c]{@{}l@{}}Retraining using gradient-based methods \\ cannot defend against coverage-driven methods.\end{tabular} & H5 &  \checkmark    \\ \hline
     \begin{tabular}[c]{@{}l@{}}Both non-differentiable and differentiable \\ transformations have similar ability in \\ improving model quality.\end{tabular} & H6 &   \\ \hline
    \end{tabular}
    \label{tab:summary}
    \end{table}

\section{Related Work}
\label{sec:related_work}
In this section, we survey some related works about DNN models, DNN testing and adversarial attack. 
Section \ref{sec:background} provides more detailed information about the test case generation methods investigated in this replication study.

\subsection{Testing and Verifying DNNs}
Models that are based on deep neural networks have achieved competitive results on many tasks, such as translation~\cite{DBLP:journals/corr/BahdanauCB14}, speech recognition~\cite{abdel2014convolutional}, semantic segmentation~\cite{CONTA},  game playing~\cite{mnih2015human} and so on. Such models are now also widely used in our daily life, e.g. autonomous driving~\cite{levinson2011towards}, ensuring the DNN quality is of great importance.

Researchers from software communities have proposed many methods to test DNN models. Inspired by using code coverage as guidance to test conventional software, Pei et al.~\cite{DeepXplore} present the concept of neuron coverage and DeepXplore to detect behaviour inconsistencies of different DNNs. The follower researchers propose more structured neuron coverage metrics and design a list of coverage-driven tools (e.g., DeepHunter~\cite{DeepHunter}, DeepTest~\cite{DeepTest}, DeepGauge~\cite{DeepGauge}, DeepCT~\cite{DeepCT}, etc.) to test DNNs. 
But recently, there are some studies questioning the effectiveness of using neuron coverage as guidance to test DNNs. Li et al.~\cite{li2019structural} point out that structural coverage criteria for neural networks could be misleading. Dong et al.~\cite{9376208} shows that there is a limited correlation between neuron coverage and robustness for DNNs. Harel-Canada et al.~\cite{correlation_fse_2020_2} show that neuron coverage may not be a useful measure for testing DNN, and Yan et al.~\cite{FSE_Yan} also find no strong correlations between neuron coverage and model quality metrics. Trujillo et al.~\cite{rl_coverage} find that it is ineffective to use neuron coverage to test deep reinforcement learning.
These studies are conducted on relatively small datasets and models and do not follow the process of how DNNs are tested and improved in practice. So we replicate the study by~\cite{FSE_Yan} and extend the experiments to more systematically compare coverage-driven methods and gradient-based methods.

There are also DNN testing works that utilize different approaches and test models beyond image classification tasks. BiasFinder~\cite{biasfinder} leverages the metamorphic relationship to uncover inconsistency of sentiment analysis systems. Sun et al.~\cite{Sun_ICSE} propose TransRepair that combines mutation with metamorphic testing to detect inconsistency bugs in machine translation systems. Gao et al.~\cite{gao_icse} adopt genetic programming to generate images that can uncover more failed test cases and better improve model robustness. Asyrofi et al.~\cite{crossasr_pp,9609154} developed a differential testing framework to test automatic speech recognition systems. DeepConcolic~\cite{DeepConcolic} proposes a white-box DL testing technique based on concolic testing.

Research also propose verification techniques to ensure the DNNs behave as expected. $AI^2$~\cite{ai_2} uses abstract interpretation to certify the safety and robustness property of DNNs. Reluplex~\cite{Reluplex} use an SMT Solver to verify the safety property of DNNs. BiasRV~\cite{biasrv} is a tool that can verify the fairness property of sentiment analysis systems at runtime, and BiasHeal~\cite{biasheal} can repair uncovered fairness issues.

\subsection{Adversarial Attack and Training}

Although Deep Neural Network (DNN) models have achieved great success on many tasks, many research works show that state-of-the-art models are vulnerable to adversarial attacks. In 2014, Szegedy et al.~\cite{42503} firstly exposed that adding a small perturbation that is not recognizable by the human eye, to the input of DNN will produce a misclassified prediction with high confidence.

According to the target model information that an attacker can access, adversarial attacks can be divided into two types: white-box and black-box. White-box attacks can access all the information of the target models, e.g. using model parameters to compute gradients. Black-box attacks mean that the attacker only knows the inputs and outputs of target models. The gradient-based methods investigated in our replication study are white-box attack.
Goodfellow et al.~\cite{FGSM} propose the first popular attack named as Fast Gradient Sign Method (FGSM), which slightly modify image pixels following the sign of gradients on images. PGD~\cite{PGD} is a first-order universal adversary attack based on FGSM~\cite{FGSM}. 
Black-box attackers like SquareAttack~\cite{ACFH2020square} and SpatialTransformation~\cite{pmlr-v97-engstrom19a} query the model for multiple times and get outputs to estimate useful information for generating adversarial examples, e.g. approximating gradients.


\section{Conclusion and Future Work}
\label{sec:conclusion}

This paper conducts a replication study of the work by Yan et al. and performs more detailed analysis. 
Our experiment results confirm that the model quality does not monotonously improve as the coverage metrics increase. We perform statistical tests and to show that coverage-improving examples is not more effective than coverage-remaining examples in terms of improving model quality.
Our experiment results also replicate the conclusion that coverage-driven methods are less effective than gradient-based methods. We test several hypotheses and further show that even coverage-driven methods are constrained to perform differentiable transformations only, the uncovered defects still cannot be repaired by adversarial training with gradient-based methods. Defensive strategies for coverage-driven methods should be further studied.

The replication package of this study are made open-source at \url{https://github.com/soarsmu/Revisiting_Neuron_Coverage}.


\section*{Acknowledgment}
This research was supported by the Singapore Ministry of Education (MOE) Academic Research Fund (AcRF) Tier 1 grant. 
We thank Wei Quan Chu for helping in the collection of the datasets for this replication study, and we are grateful that Yan et al. explain some details of their studies \cite{FSE_Yan} to us. 

\balance
\bibliographystyle{IEEEtran}
\bibliography{ref}

\begin{thebibliography}{10}
\providecommand{\url}[1]{#1}
\csname url@samestyle\endcsname
\providecommand{\newblock}{\relax}
\providecommand{\bibinfo}[2]{#2}
\providecommand{\BIBentrySTDinterwordspacing}{\spaceskip=0pt\relax}
\providecommand{\BIBentryALTinterwordstretchfactor}{4}
\providecommand{\BIBentryALTinterwordspacing}{\spaceskip=\fontdimen2\font plus
\BIBentryALTinterwordstretchfactor\fontdimen3\font minus
  \fontdimen4\font\relax}
\providecommand{\BIBforeignlanguage}[2]{{%
\expandafter\ifx\csname l@#1\endcsname\relax
\typeout{** WARNING: IEEEtran.bst: No hyphenation pattern has been}%
\typeout{** loaded for the language `#1'. Using the pattern for}%
\typeout{** the default language instead.}%
\else
\language=\csname l@#1\endcsname
\fi
#2}}
\providecommand{\BIBdecl}{\relax}
\BIBdecl

\bibitem{ciregan2012multi}
\BIBentryALTinterwordspacing
J.~Schmidhuber, U.~Meier, and D.~Ciresan, ``Multi-column deep neural networks
  for image classification,'' in \emph{2012 IEEE Conference on Computer Vision
  and Pattern Recognition (CVPR)}.\hskip 1em plus 0.5em minus 0.4em\relax Los
  Alamitos, CA, USA: IEEE Computer Society, jun 2012, pp. 3642--3649. [Online].
  Available:
  \url{https://doi.ieeecomputersociety.org/10.1109/CVPR.2012.6248110}
\BIBentrySTDinterwordspacing

\bibitem{abdel2014convolutional}
\BIBentryALTinterwordspacing
O.~Abdel-Hamid, A.-r. Mohamed, H.~Jiang, L.~Deng, G.~Penn, and D.~Yu,
  ``Convolutional neural networks for speech recognition,'' \emph{IEEE/ACM
  Transactions on Audio, Speech, and Language Processing}, vol.~22, no.~10, pp.
  1533--1545, 2014. [Online]. Available:
  \url{https://doi.org/10.1109/TASLP.2014.2339736}
\BIBentrySTDinterwordspacing

\bibitem{mnih2015human}
\BIBentryALTinterwordspacing
V.~Mnih, K.~Kavukcuoglu, D.~Silver, A.~A. Rusu, J.~Veness, M.~G. Bellemare,
  A.~Graves, M.~Riedmiller, A.~K. Fidjeland, G.~Ostrovski \emph{et~al.},
  ``Human-level control through deep reinforcement learning,'' \emph{nature},
  vol. 518, no. 7540, pp. 529--533, 2015. [Online]. Available:
  \url{https://doi.org/10.1038/nature14236}
\BIBentrySTDinterwordspacing

\bibitem{levinson2011towards}
\BIBentryALTinterwordspacing
J.~Levinson, J.~Askeland, J.~Becker, J.~Dolson, D.~Held, S.~Kammel, J.~Z.
  Kolter, D.~Langer, O.~Pink, V.~Pratt, M.~Sokolsky, G.~Stanek, D.~Stavens,
  A.~Teichman, M.~Werling, and S.~Thrun, ``Towards fully autonomous driving:
  Systems and algorithms,'' in \emph{2011 IEEE Intelligent Vehicles Symposium
  (IV)}, 2011, pp. 163--168. [Online]. Available:
  \url{https://doi.org/10.1109/IVS.2011.5940562}
\BIBentrySTDinterwordspacing

\bibitem{choi2017gram}
\BIBentryALTinterwordspacing
E.~Choi, M.~T. Bahadori, L.~Song, W.~F. Stewart, and J.~Sun, ``Gram:
  Graph-based attention model for healthcare representation learning,'' in
  \emph{Proceedings of the 23rd ACM SIGKDD International Conference on
  Knowledge Discovery and Data Mining}, ser. KDD '17.\hskip 1em plus 0.5em
  minus 0.4em\relax New York, NY, USA: Association for Computing Machinery,
  2017, p. 787–795. [Online]. Available:
  \url{https://doi.org/10.1145/3097983.3098126}
\BIBentrySTDinterwordspacing

\bibitem{gopinath2014code}
\BIBentryALTinterwordspacing
R.~Gopinath, C.~Jensen, and A.~Groce, ``Code coverage for suite evaluation by
  developers,'' in \emph{Proceedings of the 36th International Conference on
  Software Engineering}, ser. ICSE 2014.\hskip 1em plus 0.5em minus 0.4em\relax
  New York, NY, USA: Association for Computing Machinery, 2014, p. 72–82.
  [Online]. Available: \url{https://doi.org/10.1145/2568225.2568278}
\BIBentrySTDinterwordspacing

\bibitem{DeepXplore}
\BIBentryALTinterwordspacing
K.~Pei, Y.~Cao, J.~Yang, and S.~Jana, ``Deepxplore: Automated whitebox testing
  of deep learning systems,'' \emph{Commun. ACM}, vol.~62, no.~11, p.
  137–145, Oct. 2019. [Online]. Available:
  \url{https://doi.org/10.1145/3361566}
\BIBentrySTDinterwordspacing

\bibitem{DeepHunter}
\BIBentryALTinterwordspacing
X.~Xie, L.~Ma, F.~Juefei-Xu, M.~Xue, H.~Chen, Y.~Liu, J.~Zhao, B.~Li, J.~Yin,
  and S.~See, ``Deephunter: A coverage-guided fuzz testing framework for deep
  neural networks,'' in \emph{Proceedings of the 28th ACM SIGSOFT International
  Symposium on Software Testing and Analysis}, ser. ISSTA 2019.\hskip 1em plus
  0.5em minus 0.4em\relax New York, NY, USA: Association for Computing
  Machinery, 2019, p. 146–157. [Online]. Available:
  \url{https://doi.org/10.1145/3293882.3330579}
\BIBentrySTDinterwordspacing

\bibitem{DeepGauge}
\BIBentryALTinterwordspacing
L.~Ma, F.~Juefei-Xu, F.~Zhang, J.~Sun, M.~Xue, B.~Li, C.~Chen, T.~Su, L.~Li,
  Y.~Liu, J.~Zhao, and Y.~Wang, ``Deepgauge: Multi-granularity testing criteria
  for deep learning systems,'' in \emph{Proceedings of the 33rd ACM/IEEE
  International Conference on Automated Software Engineering}, ser. ASE
  2018.\hskip 1em plus 0.5em minus 0.4em\relax New York, NY, USA: Association
  for Computing Machinery, 2018, p. 120–131. [Online]. Available:
  \url{https://doi.org/10.1145/3238147.3238202}
\BIBentrySTDinterwordspacing

\bibitem{li2019structural}
\BIBentryALTinterwordspacing
Z.~Li, X.~Ma, C.~Xu, and C.~Cao, ``Structural coverage criteria for neural
  networks could be misleading,'' in \emph{2019 IEEE/ACM 41st International
  Conference on Software Engineering: New Ideas and Emerging Results
  (ICSE-NIER)}, 2019, pp. 89--92. [Online]. Available:
  \url{https://doi.org/10.1109/ICSE-NIER.2019.00031}
\BIBentrySTDinterwordspacing

\bibitem{FSE_Yan}
\BIBentryALTinterwordspacing
S.~Yan, G.~Tao, X.~Liu, J.~Zhai, S.~Ma, L.~Xu, and X.~Zhang, ``Correlations
  between deep neural network model coverage criteria and model quality,'' in
  \emph{Proceedings of the 28th ACM Joint Meeting on European Software
  Engineering Conference and Symposium on the Foundations of Software
  Engineering}, ser. ESEC/FSE 2020.\hskip 1em plus 0.5em minus 0.4em\relax New
  York, NY, USA: Association for Computing Machinery, 2020, p. 775–787.
  [Online]. Available: \url{https://doi.org/10.1145/3368089.3409671}
\BIBentrySTDinterwordspacing

\bibitem{harel2020neuron}
\BIBentryALTinterwordspacing
F.~Harel-Canada, L.~Wang, M.~A. Gulzar, Q.~Gu, and M.~Kim, ``Is neuron coverage
  a meaningful measure for testing deep neural networks?'' in \emph{Proceedings
  of the 28th ACM Joint Meeting on European Software Engineering Conference and
  Symposium on the Foundations of Software Engineering}, ser. ESEC/FSE
  2020.\hskip 1em plus 0.5em minus 0.4em\relax New York, NY, USA: Association
  for Computing Machinery, 2020, p. 851–862. [Online]. Available:
  \url{https://doi.org/10.1145/3368089.3409754}
\BIBentrySTDinterwordspacing

\bibitem{PGD}
\BIBentryALTinterwordspacing
A.~Madry, A.~Makelov, L.~Schmidt, D.~Tsipras, and A.~Vladu, ``Towards deep
  learning models resistant to adversarial attacks,'' in \emph{6th
  International Conference on Learning Representations, {ICLR} 2018, Vancouver,
  BC, Canada, April 30 - May 3, 2018, Conference Track Proceedings}.\hskip 1em
  plus 0.5em minus 0.4em\relax OpenReview.net, 2018. [Online]. Available:
  \url{https://openreview.net/forum?id=rJzIBfZAb}
\BIBentrySTDinterwordspacing

\bibitem{10.1145/1101908.1101949}
\BIBentryALTinterwordspacing
J.~A. Jones and M.~J. Harrold, ``Empirical evaluation of the tarantula
  automatic fault-localization technique,'' in \emph{Proceedings of the 20th
  IEEE/ACM International Conference on Automated Software Engineering}, ser.
  ASE '05.\hskip 1em plus 0.5em minus 0.4em\relax New York, NY, USA:
  Association for Computing Machinery, 2005, p. 273–282. [Online]. Available:
  \url{https://doi.org/10.1145/1101908.1101949}
\BIBentrySTDinterwordspacing

\bibitem{10.1145/2597008.2597148}
\BIBentryALTinterwordspacing
S.~Wang and D.~Lo, ``Version history, similar report, and structure: Putting
  them together for improved bug localization,'' in \emph{Proceedings of the
  22nd International Conference on Program Comprehension}, ser. ICPC
  2014.\hskip 1em plus 0.5em minus 0.4em\relax New York, NY, USA: Association
  for Computing Machinery, 2014, p. 53–63. [Online]. Available:
  \url{https://doi.org/10.1145/2597008.2597148}
\BIBentrySTDinterwordspacing

\bibitem{10.1109/ICSE.2017.62}
\BIBentryALTinterwordspacing
S.~Pearson, J.~Campos, R.~Just, G.~Fraser, R.~Abreu, M.~D. Ernst, D.~Pang, and
  B.~Keller, ``Evaluating and improving fault localization,'' in
  \emph{Proceedings of the 39th International Conference on Software
  Engineering}, ser. ICSE '17.\hskip 1em plus 0.5em minus 0.4em\relax IEEE
  Press, 2017, p. 609–620. [Online]. Available:
  \url{https://doi.org/10.1109/ICSE.2017.62}
\BIBentrySTDinterwordspacing

\bibitem{kim2021datasets}
M.~Kim and E.~Lee, ``Are datasets for information retrieval-based bug
  localization techniques trustworthy?'' \emph{Empirical Software Engineering},
  vol.~26, no.~3, pp. 1--66, 2021.

\bibitem{eurosat}
\BIBentryALTinterwordspacing
P.~Helber, B.~Bischke, A.~Dengel, and D.~Borth, ``Introducing eurosat: A novel
  dataset and deep learning benchmark for land use and land cover
  classification,'' in \emph{IGARSS 2018 - 2018 IEEE International Geoscience
  and Remote Sensing Symposium}, 2018, pp. 204--207. [Online]. Available:
  \url{https://doi.org/10.1109/IGARSS.2018.8519248}
\BIBentrySTDinterwordspacing

\bibitem{resnet}
\BIBentryALTinterwordspacing
K.~He, X.~Zhang, S.~Ren, and J.~Sun, ``Deep residual learning for image
  recognition,'' in \emph{2016 IEEE Conference on Computer Vision and Pattern
  Recognition (CVPR)}, 2016, pp. 770--778. [Online]. Available:
  \url{https://doi.org/10.1109/CVPR.2016.90}
\BIBentrySTDinterwordspacing

\bibitem{FGSM}
\BIBentryALTinterwordspacing
I.~Goodfellow, J.~Shlens, and C.~Szegedy, ``Explaining and harnessing
  adversarial examples,'' in \emph{International Conference on Learning
  Representations}, 2015. [Online]. Available:
  \url{http://arxiv.org/abs/1412.6572}
\BIBentrySTDinterwordspacing

\bibitem{mann1947test}
H.~B. Mann and D.~R. Whitney, ``On a test of whether one of two random
  variables is stochastically larger than the other,'' \emph{The annals of
  mathematical statistics}, pp. 50--60, 1947.

\bibitem{madry2018towards}
\BIBentryALTinterwordspacing
A.~Madry, A.~Makelov, L.~Schmidt, D.~Tsipras, and A.~Vladu, ``Towards deep
  learning models resistant to adversarial attacks,'' in \emph{International
  Conference on Learning Representations}, 2018. [Online]. Available:
  \url{https://openreview.net/forum?id=rJzIBfZAb}
\BIBentrySTDinterwordspacing

\bibitem{AdversarialTraining}
\BIBentryALTinterwordspacing
A.~Shafahi, M.~Najibi, M.~A. Ghiasi, Z.~Xu, J.~Dickerson, C.~Studer, L.~S.
  Davis, G.~Taylor, and T.~Goldstein, \emph{Adversarial training for
  free!}\hskip 1em plus 0.5em minus 0.4em\relax Curran Associates, Inc., 2019,
  vol.~32. [Online]. Available:
  \url{https://proceedings.neurips.cc/paper/2019/file/7503cfacd12053d309b6bed5c89de212-Paper.pdf}
\BIBentrySTDinterwordspacing

\bibitem{replication}
``Replication package,''
  \url{https://github.com/soarsmu/Revisiting_Neuron_Coverage.git}, accessed:
  2021-11-19.

\bibitem{SSIM}
A.~Hor\'{e} and D.~Ziou, ``Image quality metrics: Psnr vs. ssim,'' in
  \emph{2010 20th International Conference on Pattern Recognition}, 2010, pp.
  2366--2369.

\bibitem{google_coverage}
\BIBentryALTinterwordspacing
M.~Ivankovi\'{c}, G.~Petrovi\'{c}, R.~Just, and G.~Fraser, ``Code coverage at
  google,'' in \emph{Proceedings of the 2019 27th ACM Joint Meeting on European
  Software Engineering Conference and Symposium on the Foundations of Software
  Engineering}, ser. ESEC/FSE 2019.\hskip 1em plus 0.5em minus 0.4em\relax New
  York, NY, USA: Association for Computing Machinery, 2019, p. 955–963.
  [Online]. Available: \url{https://doi.org/10.1145/3338906.3340459}
\BIBentrySTDinterwordspacing

\bibitem{icse_no_correaltion}
\BIBentryALTinterwordspacing
L.~Inozemtseva and R.~Holmes, ``Coverage is not strongly correlated with test
  suite effectiveness,'' in \emph{Proceedings of the 36th International
  Conference on Software Engineering}, ser. ICSE 2014.\hskip 1em plus 0.5em
  minus 0.4em\relax New York, NY, USA: Association for Computing Machinery,
  2014, p. 435–445. [Online]. Available:
  \url{https://doi.org/10.1145/2568225.2568271}
\BIBentrySTDinterwordspacing

\bibitem{CovTesting}
``Yan et al's replication package,''
  \url{https://github.com/RU-System-Software-and-Security/CovTesting.git},
  accessed: 2021-11-19.

\bibitem{ye2021automated}
H.~Ye, M.~Martinez, and M.~Monperrus, ``Automated patch assessment for program
  repair at scale,'' \emph{Empirical Software Engineering}, vol.~26, no.~2, pp.
  1--38, 2021.

\bibitem{mnist}
\BIBentryALTinterwordspacing
L.~Deng, ``The mnist database of handwritten digit images for machine learning
  research [best of the web],'' \emph{IEEE Signal Processing Magazine},
  vol.~29, no.~6, pp. 141--142, 2012. [Online]. Available:
  \url{https://doi.org/10.1109/MSP.2012.2211477}
\BIBentrySTDinterwordspacing

\bibitem{CIFAR10}
\BIBentryALTinterwordspacing
A.~Krizhevsky, ``Learning multiple layers of features from tiny images,'' CS
  Toronto, Tech. Rep., 2009. [Online]. Available:
  \url{https://www.cs.toronto.edu/~kriz/learning-features-2009-TR.pdf}
\BIBentrySTDinterwordspacing

\bibitem{svhn}
\BIBentryALTinterwordspacing
Y.~Netzer, T.~Wang, A.~Coates, A.~Bissacco, B.~Wu, and A.~Y. Ng, ``Reading
  digits in natural images with unsupervised feature learning,'' in \emph{NIPS
  Workshop on Deep Learning and Unsupervised Feature Learning 2011}, 2011.
  [Online]. Available:
  \url{http://ufldl.stanford.edu/housenumbers/nips2011_housenumbers.pdf}
\BIBentrySTDinterwordspacing

\bibitem{DeepConcolic}
\BIBentryALTinterwordspacing
Y.~Sun, M.~Wu, W.~Ruan, X.~Huang, M.~Kwiatkowska, and D.~Kroening, ``Concolic
  testing for deep neural networks,'' in \emph{Proceedings of the 33rd ACM/IEEE
  International Conference on Automated Software Engineering}, ser. ASE
  2018.\hskip 1em plus 0.5em minus 0.4em\relax New York, NY, USA: Association
  for Computing Machinery, 2018, p. 109–119. [Online]. Available:
  \url{https://doi-org.libproxy.smu.edu.sg/10.1145/3238147.3238172}
\BIBentrySTDinterwordspacing

\bibitem{DeepGini}
\BIBentryALTinterwordspacing
Y.~Feng, Q.~Shi, X.~Gao, J.~Wan, C.~Fang, and Z.~Chen, ``Deepgini: Prioritizing
  massive tests to enhance the robustness of deep neural networks,'' in
  \emph{Proceedings of the 29th ACM SIGSOFT International Symposium on Software
  Testing and Analysis}, ser. ISSTA 2020.\hskip 1em plus 0.5em minus
  0.4em\relax New York, NY, USA: Association for Computing Machinery, 2020, p.
  177–188. [Online]. Available:
  \url{https://doi-org.libproxy.smu.edu.sg/10.1145/3395363.3397357}
\BIBentrySTDinterwordspacing

\bibitem{art}
``Adversarial robustness toolbox (art),''
  \url{https://github.com/Trusted-AI/adversarial-robustness-toolbox}, accessed:
  2021-11-19.

\bibitem{DBLP:journals/corr/BahdanauCB14}
\BIBentryALTinterwordspacing
D.~Bahdanau, K.~Cho, and Y.~Bengio, ``Neural machine translation by jointly
  learning to align and translate,'' in \emph{3rd International Conference on
  Learning Representations, {ICLR} 2015, San Diego, CA, USA, May 7-9, 2015,
  Conference Track Proceedings}, Y.~Bengio and Y.~LeCun, Eds., 2015. [Online].
  Available: \url{http://arxiv.org/abs/1409.0473}
\BIBentrySTDinterwordspacing

\bibitem{CONTA}
\BIBentryALTinterwordspacing
D.~Zhang, H.~Zhang, J.~Tang, X.-S. Hua, and Q.~Sun, ``Causal intervention for
  weakly-supervised semantic segmentation,'' in \emph{Advances in Neural
  Information Processing Systems}, H.~Larochelle, M.~Ranzato, R.~Hadsell, M.~F.
  Balcan, and H.~Lin, Eds., vol.~33.\hskip 1em plus 0.5em minus 0.4em\relax
  Curran Associates, Inc., 2020, pp. 655--666. [Online]. Available:
  \url{https://proceedings.neurips.cc/paper/2020/file/07211688a0869d995947a8fb11b215d6-Paper.pdf}
\BIBentrySTDinterwordspacing

\bibitem{DeepTest}
\BIBentryALTinterwordspacing
Y.~Tian, K.~Pei, S.~Jana, and B.~Ray, ``Deeptest: Automated testing of
  deep-neural-network-driven autonomous cars,'' in \emph{Proceedings of the
  40th International Conference on Software Engineering}, ser. ICSE '18.\hskip
  1em plus 0.5em minus 0.4em\relax New York, NY, USA: Association for Computing
  Machinery, 2018, p. 303–314. [Online]. Available:
  \url{https://doi.org/10.1145/3180155.3180220}
\BIBentrySTDinterwordspacing

\bibitem{DeepCT}
\BIBentryALTinterwordspacing
L.~Ma, F.~Juefei-Xu, M.~Xue, B.~Li, L.~Li, Y.~Liu, and J.~Zhao, ``Deepct:
  Tomographic combinatorial testing for deep learning systems,'' in \emph{2019
  IEEE 26th International Conference on Software Analysis, Evolution and
  Reengineering (SANER)}, Feb 2019, pp. 614--618. [Online]. Available:
  \url{https://doi.org/10.1109/SANER.2019.8668044}
\BIBentrySTDinterwordspacing

\bibitem{9376208}
\BIBentryALTinterwordspacing
Y.~Dong, P.~Zhang, J.~Wang, S.~Liu, J.~Sun, J.~Hao, X.~Wang, L.~Wang, J.~Dong,
  and T.~Dai, ``An empirical study on correlation between coverage and
  robustness for deep neural networks,'' in \emph{2020 25th International
  Conference on Engineering of Complex Computer Systems (ICECCS)}, 2020, pp.
  73--82. [Online]. Available:
  \url{https://doi.org/10.1109/ICECCS51672.2020.00016}
\BIBentrySTDinterwordspacing

\bibitem{correlation_fse_2020_2}
\BIBentryALTinterwordspacing
F.~Harel-Canada, L.~Wang, M.~A. Gulzar, Q.~Gu, and M.~Kim, ``Is neuron coverage
  a meaningful measure for testing deep neural networks?'' in \emph{Proceedings
  of the 28th ACM Joint Meeting on European Software Engineering Conference and
  Symposium on the Foundations of Software Engineering}, ser. ESEC/FSE
  2020.\hskip 1em plus 0.5em minus 0.4em\relax New York, NY, USA: Association
  for Computing Machinery, 2020, p. 851–862. [Online]. Available:
  \url{https://doi.org/10.1145/3368089.3409754}
\BIBentrySTDinterwordspacing

\bibitem{rl_coverage}
\BIBentryALTinterwordspacing
M.~Trujillo, M.~Linares-V\'{a}squez, C.~Escobar-Vel\'{a}squez, I.~Dusparic, and
  N.~Cardozo, ``Does neuron coverage matter for deep reinforcement learning? a
  preliminary study,'' in \emph{Proceedings of the IEEE/ACM 42nd International
  Conference on Software Engineering Workshops}, ser. ICSEW'20.\hskip 1em plus
  0.5em minus 0.4em\relax New York, NY, USA: Association for Computing
  Machinery, 2020, p. 215–220. [Online]. Available:
  \url{https://doi.org/10.1145/3387940.3391462}
\BIBentrySTDinterwordspacing

\bibitem{biasfinder}
M.~H. Asyrofi, Z.~Yang, I.~N.~B. Yusuf, H.~J. Kang, F.~Thung, and D.~Lo,
  ``Biasfinder: Metamorphic test generation to uncover bias for sentiment
  analysis systems,'' \emph{IEEE Transactions on Software Engineering}, 2021.

\bibitem{Sun_ICSE}
\BIBentryALTinterwordspacing
Z.~Sun, J.~M. Zhang, M.~Harman, M.~Papadakis, and L.~Zhang, ``Automatic testing
  and improvement of machine translation,'' in \emph{Proceedings of the
  ACM/IEEE 42nd International Conference on Software Engineering}, ser. ICSE
  '20.\hskip 1em plus 0.5em minus 0.4em\relax New York, NY, USA: Association
  for Computing Machinery, 2020, p. 974–985. [Online]. Available:
  \url{https://doi.org/10.1145/3377811.3380420}
\BIBentrySTDinterwordspacing

\bibitem{gao_icse}
\BIBentryALTinterwordspacing
X.~Gao, R.~K. Saha, M.~R. Prasad, and A.~Roychoudhury, ``Fuzz testing based
  data augmentation to improve robustness of deep neural networks,'' in
  \emph{Proceedings of the ACM/IEEE 42nd International Conference on Software
  Engineering}, ser. ICSE '20.\hskip 1em plus 0.5em minus 0.4em\relax New York,
  NY, USA: Association for Computing Machinery, 2020, p. 1147–1158. [Online].
  Available: \url{https://doi-org.libproxy.smu.edu.sg/10.1145/3377811.3380415}
\BIBentrySTDinterwordspacing

\bibitem{crossasr_pp}
\BIBentryALTinterwordspacing
M.~H. Asyrofi, Z.~Yang, and D.~Lo, ``Crossasr++: A modular differential testing
  framework for automatic speech recognition,'' in \emph{Proceedings of the
  29th ACM Joint Meeting on European Software Engineering Conference and
  Symposium on the Foundations of Software Engineering}, ser. ESEC/FSE
  2021.\hskip 1em plus 0.5em minus 0.4em\relax New York, NY, USA: Association
  for Computing Machinery, 2021, p. 1575–1579. [Online]. Available:
  \url{https://doi-org.libproxy.smu.edu.sg/10.1145/3468264.3473124}
\BIBentrySTDinterwordspacing

\bibitem{9609154}
M.~H. Asyrofi, Z.~Yang, J.~Shi, C.~W. Quan, and D.~Lo, ``Can differential
  testing improve automatic speech recognition systems?'' in \emph{2021 IEEE
  International Conference on Software Maintenance and Evolution (ICSME)},
  2021, pp. 674--678.

\bibitem{ai_2}
\BIBentryALTinterwordspacing
T.~Gehr, M.~Mirman, D.~Drachsler-Cohen, P.~Tsankov, S.~Chaudhuri, and
  M.~Vechev, ``Ai2: Safety and robustness certification of neural networks with
  abstract interpretation,'' in \emph{2018 IEEE Symposium on Security and
  Privacy (SP)}, May 2018, pp. 3--18. [Online]. Available:
  \url{https://doi.org/10.1109/SP.2018.00058}
\BIBentrySTDinterwordspacing

\bibitem{Reluplex}
\BIBentryALTinterwordspacing
G.~Katz, C.~Barrett, D.~L. Dill, K.~Julian, and M.~J. Kochenderfer, ``Reluplex:
  An efficient smt solver for verifying deep neural networks,'' in
  \emph{Computer Aided Verification}, R.~Majumdar and V.~Kun{\v{c}}ak,
  Eds.\hskip 1em plus 0.5em minus 0.4em\relax Cham: Springer International
  Publishing, 2017, pp. 97--117. [Online]. Available:
  \url{https://link.springer.com/chapter/10.1007/978-3-319-63387-9_5}
\BIBentrySTDinterwordspacing

\bibitem{biasrv}
\BIBentryALTinterwordspacing
Z.~Yang, M.~H. Asyrofi, and D.~Lo, ``Biasrv: Uncovering biased sentiment
  predictions at runtime,'' in \emph{Proceedings of the 29th ACM Joint Meeting
  on European Software Engineering Conference and Symposium on the Foundations
  of Software Engineering}, ser. ESEC/FSE 2021.\hskip 1em plus 0.5em minus
  0.4em\relax New York, NY, USA: Association for Computing Machinery, 2021, p.
  1540–1544. [Online]. Available:
  \url{https://doi.org/10.1145/3468264.3473117}
\BIBentrySTDinterwordspacing

\bibitem{biasheal}
Z.~Yang, H.~Jain, J.~Shi, M.~H. Asyrofi, and D.~Lo, ``Biasheal: On-the-fly
  black-box healing of bias in sentiment analysis systems,'' in \emph{2021 IEEE
  International Conference on Software Maintenance and Evolution (ICSME)},
  2021, pp. 644--648.

\bibitem{42503}
\BIBentryALTinterwordspacing
C.~Szegedy, W.~Zaremba, I.~Sutskever, J.~Bruna, D.~Erhan, I.~J. Goodfellow, and
  R.~Fergus, ``Intriguing properties of neural networks,'' in \emph{2nd
  International Conference on Learning Representations, {ICLR} 2014, Banff, AB,
  Canada, April 14-16, 2014, Conference Track Proceedings}, Y.~Bengio and
  Y.~LeCun, Eds., 2014. [Online]. Available:
  \url{http://arxiv.org/abs/1312.6199}
\BIBentrySTDinterwordspacing

\bibitem{ACFH2020square}
\BIBentryALTinterwordspacing
M.~Andriushchenko, F.~Croce, N.~Flammarion, and M.~Hein, ``Square attack: a
  query-efficient black-box adversarial attack via random search,'' in
  \emph{European Conference on Computer Vision (ECCV)}, 2020. [Online].
  Available: \url{https://arxiv.org/abs/1912.00049}
\BIBentrySTDinterwordspacing

\bibitem{pmlr-v97-engstrom19a}
\BIBentryALTinterwordspacing
L.~Engstrom, B.~Tran, D.~Tsipras, L.~Schmidt, and A.~Madry, ``Exploring the
  landscape of spatial robustness,'' in \emph{Proceedings of the 36th
  International Conference on Machine Learning}, ser. Proceedings of Machine
  Learning Research, K.~Chaudhuri and R.~Salakhutdinov, Eds., vol.~97.\hskip
  1em plus 0.5em minus 0.4em\relax Long Beach, California, USA: PMLR, 09--15
  Jun 2019, pp. 1802--1811. [Online]. Available:
  \url{http://proceedings.mlr.press/v97/engstrom19a.html}
\BIBentrySTDinterwordspacing

\end{thebibliography}

\end{document}